\documentstyle[aps,prl,multicol]{revtex}
\begin{document}
\draft 
\title{ Is there a Phase Transition to the Flux Lattice State?}
\author{M.A.Moore}
\address{Department of Physics, University of Manchester,
Manchester, M13 9PL, United Kingdom.}
\date{\today}
\maketitle

\begin{abstract}
The sharp drops in the resistance and magnetization which are usually 
attributed to a phase transition from the vortex liquid state to a  crystal 
state are explained instead as a crossover  between three and two 
dimensional behavior, which occurs when the phase coherence length 
 in the liquid becomes comparable to the sample 
thickness.  Estimates of the width of the crossover region and the phase 
coherence length scales are in  agreement with experiment.  
\end{abstract} 

%\narrowtext
\begin{multicols}{2}
%%%%%%%%%%%%
For some years now it has been widely believed that  the vortex liquid state 
of a type-II superconductor changes on cooling  to a flux lattice 
 at a first-order phase transition \cite{Blatter} .  Sharp drops have been 
seen in the magnetization \cite {Zeldov,Welp,Hardy}, entropy \cite 
{Schilling} and resistance \cite{Safar}  
 and it is usually assumed that such drops are  due to a  freezing 
 transition. I shall describe here a simple alternative explanation   of 
 these drops, based on the idea that they are caused by a crossover from 
 three-dimensional 
behavior to two-dimensional behavior when the phase correlation length  
along the field direction in the vortex liquid,  $ l_ \|$, becomes 
comparable to the sample dimensions \cite{foot}.  
 Because  $l_\|$ grows very rapidly 
as the temperature is lowered the crossover region appears narrow  
enough to masquerade as a first order phase transition. An estimate of its 
 width is given  and  found to be comparable with the width of the 
drop  in the magnetization  \cite{Welp}.

Central to my picture is the idea that at all non-zero temperatures the 
system is in the vortex liquid state and that it is only finite size 
effects which produce the semblance of a phase transition.  In the 
thermodynamic limit of an infinitely big system there would be no phase
 transition, 
no sharp drops and only a vortex liquid phase. At zero temperature in 
the ground state there would of course be the usual Abrikosov lattice.  The 
length scale over which there is short-range crystalline order in the 
vortex liquid, $l_\perp$, is expected to increase without limit as the 
temperature is lowered towards absolute zero.  

Essentially  the picture being advocated here is that of Refs. 
\cite{moore,Moore}, where it was shown that in both two and three 
dimensions phase coherence in the vortex crystal state was 
destroyed  by the phase fluctuations associated with the shear modes of the 
crystal  and it was suggested  that as a consequence crystalline order 
might not exist in these dimensions. Since it is not  obvious why the 
loss of phase coherence need necessarily imply the absence of crystalline 
order this suggestion  has (to say the least) not been widely accepted. 
However, our Monte Carlo simulations in two-dimensions \cite{om,hanlee,dm}, 
 done within the lowest Landau level approximation 
(LLL), found no finite temperature phase transition and demonstrated that 
the length scale over which phase coherence is lost is  $l_\perp$, which 
diverged  in the zero-temperature limit.  

Strong support for the idea that the apparent first-order phase transition  
is just a crossover from three to two-dimensional behavior comes from  
 multi-terminal transport (flux transformer) measurements on 
untwinned single crystals of YBCO \cite{Lopez} and on thin  BSCCO crystals 
\cite{Garland}. In these experiments current is injected and removed at two 
electrodes attached to the top face of the crystal and  voltage 
differences  measured  between two points lying between the electrodes 
 on the top face, $V_{TOP}$, and between two  points directly below them on 
the bottom face of the crystal, $V_{BOT}$. The magnetic field is along the 
c-axis of the crystal and
the thicknesses  of the YBCO crystals ranged up to $ 35 \ \mu m$ while the 
BSCCO crystals were  around $1 \ \mu m$ thick.  For  crystals of these 
thicknesses both voltages became very nearly equal to each other as the 
temperature was reduced after which both  fell sharply to zero. The 
vanishing of the voltage is usually taken as marking the onset of the 
freezing transition into the crystalline state.  When $V_{TOP}$ and 
$V_{BOT}$ are equal it means the flux lines are moving  without cutting and 
reconnecting so there must be  phase coherence across the crystal. As 
$V_{TOP}$ and $V_{BOT}$ are becoming equal  as the \lq\lq freezing" 
transition is approached, then the length scale over which there is phase 
coherence in the direction of the magnetic field, $l_\|$, must be growing 
as the temperature of the vortex liquid is reduced,   becoming comparable 
to the crystal thickness right at the apparent phase transition. 
Furthermore, measurements of the non-linear current-voltage characteristics 
on the same crystals give information on the length scale $l_\perp$ via the 
scaling relation  $\sigma E=Jf(Jl_\perp l_\| \Phi_0/k_BT)$,  where $E$ is 
the electric field perpendicular to the field, $J$ is the current density 
in the same direction, $\sigma$ is the conductivity and $\Phi_0$ is the 
flux quantum \cite{FFH}. Preliminary studies indicate that it too is a 
length 
of micron magnitude \cite{Righi}. In the conventional picture of  a first-
order phase transition it is very hard to understand why long-length scales 
should be present in the vortex liquid --- $l_\|$ is at least as large as 
$35  \ \mu m$ in YBCO at the freezing transition ---and why there should be 
a  coincidence of the freezing transition with the temperature at which 
this length scale becomes comparable to the dimensions of the sample. My 
hypothesis that the \lq \lq freezing" phenomenon is a crossover induced 
effect seems much more natural.

We start from the Ginzburg-Landau free 
energy for a superconductor in a uniform magnetic field ${\bf B}={\bf\nabla}
\times{\bf A}$;
\begin {eqnarray} 
F[\Psi ]&=& \int d^3{\bf 
r}\bigg(\alpha(T) |\Psi |^2 +\frac{\beta}{2} |\Psi |^4 \nonumber \\
&&+\ \sum_\mu 
\frac{1}{2m_\mu }|(-i\hbar\partial_\mu - 2e A_\mu)\Psi |^2\bigg).  
\label{LG} 
\end{eqnarray} 
Here $ \alpha(T)$ is a temperature-dependent 
variable e.g $ \alpha(T) = -\alpha'(1- T/T_{c0})$, $T_{c0}$ is the mean-
field transition temperature, $ \beta$ is the coupling constant and $m_\mu$ 
is an effective mass.  The masses in the $ab$ plane will  be taken  equal 
and denoted by $m_{ab}$ and the mass in the c direction is  $m_c$. To start 
with we shall neglect the fluctuations in the vector potential ${\bf A}$ 
and restrict the order parameter $\Psi$ to the space spanned by the lowest 
Landau levels  (LLL).  Then  $\alpha_H\equiv\alpha(T)+ e\hbar B/m_{ab}=0$ 
defines  the mean-field  line $H_{c2}$. The temperature is conveniently and 
conventionally \cite{rt}  represented by the dimensionless parameter 
$\alpha_T\equiv\alpha_H(4\pi\hbar^2/\beta eBk_BT\sqrt{2m_c})^{2/3}$. Low 
temperature is represented by $\alpha_T \rightarrow-\infty$, high 
temperature by $\alpha_T\rightarrow\infty$. More conveniently $ 
\alpha_T=(2/Gi)^{1/3}(ht)^{-2/3}(1-t-h)$ where $Gi$ is the Ginzburg number, 
(about $0.02$ for YBCO \cite{Blatter}), $t$ is the reduced temperature 
$T/T_{c0}$ and $h=B/B_{c2}(0)$ where $B_{c2}(0)$ is the straight line 
extrapolant of the $H_{c2}$ to zero temperature (about $ 168 \ T$ in YBCO 
\cite{Blatter}).  

The length scale  $l_\|$ is a 
measure of the phase correlation length along the field direction (and  is 
technically the length scale on which the propagator 
$\langle\Psi^*(x,y,z)\Psi(x,y,z')\rangle$ decreases as a function of $|z-
z'|$. The other important length scale in the problem, $l_\perp$, which 
measures the degree of short-range crystalline order in the vortex liquid 
can be obtained from  the density-density correlation function 
$\langle|\Psi(x,y,z)|^2|\Psi(x',y',z)|^2\rangle_c$. In addition $l_\perp$ 
seems to be equal to an (appropriately defined) phase coherence length 
scale in the direction perpendicular to the field \cite{dm}. It is our 
contention that these lengths reach  infinity as the temperature goes to 
zero and  that as a consequence it is useful to use the techniques of 
\lq\lq zero-temperature scaling" \cite{bm}.  

At zero-temperature the ground state order parameter $\Psi_0(x,y)$ which 
minimises $F[\Psi]$ corresponds to the Abrikosov flux lattice of straight 
vortex lines arranged in the form of a triangular lattice.  Expanding about 
the minimum by setting $\Psi = \Psi_0+\delta\Psi[x,y,z]$ one finds 
\cite{moore,Eilen} that the low energy states above the minimum can be 
described by an effective Hamiltonian 
\begin{equation}
H=\frac{1}{2}\int d^3{\bf r}\bigg(\rho_s(\partial\theta/\partial z)^2 + 
c_{66}l^4( \nabla_\perp^2\theta)^2\bigg).  \label{etm} 
\end{equation} 
Here $\theta$ is the phase change associated with $\delta\Psi$, $\rho_s$ is 
the superfluid density ($\rho_s=\tilde{\rho}_s\hbar^2|\alpha_H|/m_c\beta$ 
in the LLL mean-field limit, with $\tilde{\rho_s}=1/\beta_A$ and 
$\beta_A=1.1596 \ldots$) and $c_{66}$ is the shear modulus equal in LLL 
mean-field approximation to $2\tilde{c}_{66} \alpha_H^2/\beta $ with 
$\tilde{c}_{66} =0.0885 \ldots$\ .  
 The magnetic length scale $l=\sqrt{\hbar/2eB}$.  The relation between the 
 phase change $\theta$ and  flux line displacements is $u_x=-
 l^2\partial\theta/\partial y$ and $u_y=l^2\partial\theta/\partial x$ 
\cite{moore}. I believe that vortex crystals behave differently from other 
crystals because only for them can  displacements be written as  phase 
derivatives.  

It is convenient to intr
oduce the dimensionless  variables $Z$  
where $z=\xi_cZ$; $\xi_c\equiv(\hbar^2/2m_c|\alpha_H|)^{1/2}$ and 
${\bf R}=(X,Y)$ where $x=lX$, $y=lY$.
Then the effective Hamiltonian in $d$ dimensions becomes
\begin{equation}
\frac{H}{k_BT}=\frac{1}{2\tilde{T}}\int d^2 {\bf R}\int d^{d-2}{\bf Z}\bigg(
\tilde{\rho}_s(\partial\theta/\partial Z_i)^2+\tilde{c}_{66}
(\tilde{\nabla}_\perp^2\theta)^2\bigg)
\label{retm}
\end{equation}
The subscript $i$ takes values $3,4, \ldots,d$ and an implied summation 
convention is being used.  In three dimensions the dimensionless effective 
temperature $\tilde{T}$ equals $4\pi/|\alpha_T|^{3/2}$.  

For the case of a thin film of thickness $L{_z}$ where the variation of 
$\Psi$ with respect to $z$ can be ignored, the Hamiltonian reduces to 
\begin{equation}
\frac{H}{k_BT}=\frac{1}{2\tilde{T}}\int d^2{\bf R}\, 
\tilde{c}_{66} (\tilde{\nabla}_\perp^2\theta)^2,
\label{2retm} 
\end{equation}
where in two dimensions the  effective temperature 
$\tilde{T}$ equals $\pi/\alpha_{2T}^{2}$, with $\alpha_{2T}=\alpha_H(\pi\hbar 
L{_z}/\beta eBk_BT)^{1/2}$.  
  
Hamiltonians such as  Eq.~(\ref{retm}) have been  studied in 
connection with smectic liquid crystals \cite{Toner} where a 
renormalisation group treatment has proved very useful. To this end rescale 
lengths so that $x'=bx, y'=by$ but $z'=b^2z$ when in order to keep the free 
energy invariant the renormalised temperature $\tilde{T}'=b^{6-
2d}\tilde{T}$. Setting $b=e^{\rho}$  one obtains the following flow 
equation for the effective temperature as a function of the length scale: 
\begin{equation} \frac{d\tilde{T}}{d\rho}=(6-2d)\tilde{T}.  \label{rgf} 
\end{equation} For dimensions $d\leq 3$ the flow takes one away from low 
temperatures to the high-temperature sink. Above three dimensions 
$\tilde{T}=0$ is a stable 
 fixed point indicating that $3$ is the lower critical dimension for  
 behavior controlled by this fixed point.  

Consider first the case of thin films ($d=2$). Starting at a low effective 
temperature $\tilde{T}$ where the correlation length is $l_\perp$, scale up 
until the effective temperature  $\tilde{T'}$ is  $a$, of order unity, and 
where the correlation length is $l$. Then $b=l_\perp/l$ and   
$a=\tilde{T'}=(l_\perp/l)^2\tilde{T}$  so $l_\perp \sim 
l(1/\tilde{T})^{1/2}\sim l|\alpha_{2T}|$.  This result for $l_{\perp}$ is 
consistent with our Monte Carlo simulations in two dimensions 
\cite{om,hanlee,dm}  and has also been derived within the so-called parquet 
graph approximation \cite{ym}.  There have been a number of other Monte 
Carlo simulations in two-dimensions employing (the experimentally 
unrealisable) quasi-periodic boundary conditions and the LLL approximation 
in which a first order transition to a crystalline state seems to 
take place at a 
finite temperature \cite{sim}. As no  first-order transition has been 
reported for superconducting films, I shall take the apparent phase 
transition to be an artifact of the simulations. For a 
review of this controversy, see Ref. \cite{dm}.  

In thin superconducting films  ac resistivities \cite{berghuis} and 
flux liquid viscosities \cite{Kesv} have been 
apparently successfully interpreted in terms of the continuous phase 
transitions predicted in conventional two-dimensional melting theory 
\cite{nelson}, which would seem to be at odds with the idea that the  
 vortex liquid state should be the only state present at all finite 
 temperatures. These experiments essentially determine the relaxational 
 timescale of the system which grows rapidly at low temperatures. According 
 to our Monte Carlo simulations  \cite{hanlee,AKK} it increases with an 
 Arrhenius form in the two-dimensional effective temperature, i.e. as 
 $~\exp(C/\tilde{T})$, with C a constant of order one. Such an expression 
 provides an adequate fit to the experimental data. Invoking a finite 
 temperature phase transition thus seems unnecessary.  

Returning now to bulk systems one sees that the term on the right-hand side 
of Eq.~(\ref{rgf}) is zero when $d=3$. To describe three dimensions it is 
necessary to go beyond the simple scaling flow equation. At low 
temperatures one expects it to become instead \cite{FN}, $d\tilde{T}/d\rho 
=\tilde{T}^2/(2\pi A)$ where A is a constant. On integrating up to the lengthscales at which $\tilde{T'}$ is of 
order unity, where $b=l_\perp/l=(l_\|/\xi_c)^{1/2}$, one has $l_\|\sim 
\xi_c \,\exp(A|\alpha_T|^{3/2})$ and $l_\perp\sim l 
\,\exp(A|\alpha_T|^{3/2}/2)$.  Such exponentially rapid growth of 
correlation lengths is typical of a system at its lower critical dimension. 
It is perhaps  noteworthy that phase coherence  length scales  of this form 
were previously 
 obtained via estimates of the crossing energy of flux lines \cite{wm2}.

When the sample thickness $L_z$ is greater than $l_\|$ then the system 
behaves as a three-dimensional bulk system, but as the temperature is 
lowered $l_\|$  grows and eventually becomes greater than $L_z$ . Then the 
vortices behave as in a two-dimensional thin-film system with phase 
coherence across the sample.  The crossover region is where $L_z\sim l_\|$. 
If one knew $A$ one could get a good estimate of  $\alpha_T^{*}$, the value 
of $\alpha_T$ at the crossover, for a  sample of known thickness .  At this 
time we shall  take $\alpha_T^{*}$  from experiment \cite{cm}. It is around 
$-9$  for YBCO crystals of (say) $40 \ \mu m$ dimensions \cite{wm1}.  It 
follows from the definition of $\alpha_T$ that the crossover line obeys the 
LLL scaling relation $B\sim (T_{c0}-T)^{3/2}$ for fixed $L_z$, except for 
temperatures so close to $T_{c0}$ that it is no longer possible to regard  
$\xi_c$ as a constant.  

 Because of the exponentially rapid growth of $l_\|$, the width of the 
crossover region as a function of field and temperature is quite narrow. 
Its width  may be  estimated by finding the small change $\delta\alpha_T$ 
in $\alpha_T^{*}$ to make $l_\|$ grow to $2L_z$. Then 
$3\delta\alpha_T/2\alpha_T^{*}\sim \ln(2)/\ln(L_z/\xi_c)\sim \delta B/B$, 
where $\delta B$ is the change in $B$ required to double $l_\|$. For the   
$0.2 \ mm$ thick crystal of Welp {\it et al.} \cite{Welp}, we predict that 
$\delta B/B$ is about $0.054$ at a field of $4.2 \ T$ while the observed 
width looks to me to be around $0.2 \ T$, making the ratio $\delta B/B$  
about $0.05$. Thus the crossover width is similar to the rounding of the  
jump in the magnetization. If one believed in a first-order transition one 
would have to claim that the rounding of the jump was somehow caused by  a 
spatial inhomogeneity of oxygen concentration or some  non-uniformity of 
the applied field. Another feature of our crossover argument is that as the 
sample thickness is increased then the crossover  should occur at lower 
temperatures, but the decrease of the \lq\lq transition" temperature  is 
only as the logarithm of the sample thickness. The magnitude of the effect 
is easily found from solving $L_z\sim l_\|$  and  certainly looks 
measurable  provided that a range of crystals can be made of various 
thicknesses but otherwise identical.  

When the crossover takes place the effective temperature of the resulting 
two-dimensional (thin film) state is very low. Using the definition of 
$\alpha_{2T}$  to re-express it in terms of $h$ and $t$ as   
$\alpha_{2T}=0.5(Gi/2)^{1/12}(ht)^{1/6}\alpha_T(L_z/\xi_0)^{1/2}$, where 
$\xi_0$ is the zero-temperature c-axis correlation length, 
$(\hbar^2/2m_c\alpha')^{1/2}$ (which is about $2 \ \AA$) and then inserting 
numbers appropriate to YBCO at a field of $4 \ T$ and sample thickness $0.2 
\ mm$ one finds $\alpha_{2T}\sim -1600$. Notice that in the crossover 
region and below,  $l_\perp$  is more than $10^3 \ l$. It would then only 
require a tiny amount of coupling between the vortex system and the 
underlying real crystal to produce a system whose structure factor 
displayed Bragg spots rather than the rings expected for a liquid state 
\cite{ym3}.  Thus the vortex system would appear  crystalline to most 
experimental probes.  

Because the two-dimensional state below the crossover is at such a low 
effective temperature, its entropy can be obtained by differentiating the 
mean-field free-energy per unit volume, $-\alpha_H^2/2\beta\beta_A$, with 
respect to the temperature. However, the entropy of the three-dimensional 
state when $\alpha_T=\alpha_T^{*}\sim -9$ is a few percent above the   
mean-field value \cite{ss3}. Therefore the entropy must decrease rapidly on 
cooling through the crossover region, giving rise to an apparent jump. Per 
vortex per double layer the entropy change $\Delta S$ can be written as 
\begin{equation}
\Delta S \sim k_B (s/\xi_0)(2/Gi)^{1/6}t^{2/3}h^{-1/3}g(\alpha_T),
\label{jump}
\end{equation}
where $h,t$ and $\alpha_T$ are at their crossover values, 
$s$ is the distance between the double layers and the function 
$g(\alpha_T)$ has at present to be inferred from the numerical simulations 
\cite{ss3}  of specific heats. I  estimate that $\Delta S$  is  around  
$k_B$  at $4\ T$.  Notice that Eq.~(\ref{jump}) predicts that $\Delta S$ 
increases as the field is reduced in accord with experiment  
\cite{Zeldov,Hardy}.  

The argument of this paper is that the properties of the vortex liquid are 
controlled by a particular zero-temperature fixed point. Because in 
dimensions two and three there are diverging length scales associated with 
this fixed point, the concept of renormalisation group relevance and 
irrelevance can be used to determine the importance of effects ignored in 
the calculation. Thus higher Landau levels are clearly irrelevant in the 
low-temperature limit as they are associated with more massive fields. 
Similarly screening effects --- the fluctuations of the vector potential 
${\bf A}$ --- can be shown to be  irrelevant at this zero-temperature fixed 
point by an extension of the argument used in Ref.\cite{BNT}, where such 
fluctuations were shown to be also irrelevant at the critical  fixed point. 
Thus our treatment of the length scales $l_\|$ and $l_\perp$ should become 
exact in the low-temperature limit. However, if there are terms present 
which couple the vortices to the underlying crystal (as in numerical 
simulations done on a grid) then a  spectrum of low-energy excitations may 
arise different from  that in Eq.~(\ref{etm})  which  could result in a 
finite temperature phase transition.  

One important effect so far neglected is disorder. After electron 
irradiation, which induces point defects, the \lq\lq drops" in the 
resistance are replaced by a smooth decrease with temperature 
\cite{Fendrich}. Disorder might be expected to limit the growth of 
$l_\perp$ and $l_\|$ at the Larkin length scales $R_\perp$ and $R_\|$  
\cite{Blatter}, which would naturally prevent a sharp crossover effect from 
occurring. Even in non-irradiated samples  the resistance drop is much 
narrower as a function of temperature  than the drop in the magnetization.  
I believe that the sharper drop  in the resistance is a consequence of a 
percolative mechanism in which regions of transverse dimensions $l_\perp$ 
and longitudinal size $l_\|$ become pinned by the disorder and when these 
pinned regions percolate across the system then the resistance falls away.  
In thicker ($90 \  \mu m$)  untwinned crystals of YBCO $V_{TOP}$ does not 
equal $V_{BOT}$ when the resistance drops to zero \cite{Lopez3} --- a fact 
consistent with a percolative mechanism.  

Thus it seems possible to give an account of the main features of the mixed 
phase at a semi-quantitative level by employing the ideas of        zero-
temperature scaling. Should there actually be a finite temperature phase 
transition then these arguments become irrelevant. However, it is hard to 
see how the long length scales revealed by the flux transformer experiments 
can be incorporated into the conventional melting picture.

% %%%%%%%%%%%%%%%%%%
 
\end{multicols}  

\begin {references}
\bibitem{foot} This crossover should not be confused with the
 \lq\lq dimensional" crossover which takes place if the flux lines decompose
  into pancakes  \cite{Blatter}.
\end{references}
\end{document}